\documentclass[a4paper]{jpconf}
\usepackage{graphicx}
\usepackage{amssymb}
\def\de{\partial}

\def\2{\frac12}
\def\4{\frac14}
\def\ie{{\it i.e.}~}
\def\eg{{\it e.g.}~}

\newcommand{\be}{\begin{equation}}
\newcommand{\ee}{\end{equation}}
\newcommand{\bea}{\begin{eqnarray}}
\newcommand{\eea}{\end{eqnarray}}
\newcommand{\ba}{\begin{array}}
\newcommand{\ea}{\end{array}}

\def\d{\delta}
\def\e{\epsilon}

\def\l{\lambda}

\def\m{\mu}

\def\de{\partial}

\begin{document}
\title{Massive higher spins and holography\footnote{Talk
presented by M.B. at the Fourth Meeting on Constrained Dynamics and
Quantum gravity held in Cala Gonone (Sardinia, Italy), September
12-16, 2005.}}

\author{Massimo Bianchi$^1$ and Fabio Riccioni$^{1,2}$}

\address{$^1$Dipartimento di Fisica, Universit{\`a} di Roma ``Tor
Vergata'', I.N.F.N. - Sezione di Roma II ``Tor Vergata'', Via della
Ricerca Scientifica, 1 - 00133 Roma - ITALY\\
$^2$DAMTP, Centre for
Mathematical Sciences, University of Cambridge, Wilberforce Road,
Cambridge CB3 0WA,  UK }

\ead{Massimo.Bianchi@roma2.infn.it, F.Riccioni@damtp.cam.ac.uk}

\begin{abstract}
We review recent progress towards the understanding of higher
spin gauge symmetry breaking in AdS space from a holographic
vantage point. According to
the AdS/CFT correspondence, ${\cal N}=4$
SYM theory at vanishing coupling constant should be dual to a theory in AdS
which exhibits higher spin gauge symmetry enhancement. When the SYM
coupling is non-zero, all but a handful of HS currents are violated by
anomalies, and correspondingly local higher spin symmetry in the
bulk gets spontaneously broken. In agreement with previous results
and holographic expectations, we find that, barring one notable exception
(spin 1 eating spin 0), the Goldstone modes
responsible for HS symmetry breaking in AdS have non-vanishing
mass even in the limit in which the gauge symmetry is restored. We
show that spontaneous breaking \'a la St\"uckelberg implies that
the mass of the relevant spin $s'=s-1$ Goldstone field is exactly the one
predicted by the correspondence.
\end{abstract}

\section{Introduction}
The AdS/CFT correspondence~\cite{adscft} between IIB superstring
theory on $AdS_5 \times S^5$ with $N$ units of 5-form flux and
$SU(N)$ ${\cal N}=4$ SYM theory in $d=4$ is most often discussed in
the limit of large AdS radius.  In this limit the AdS side is under
control, since the higher spin fields become extremely massive and
decouple from the (super)gravity modes. Instead, in the CFT side
this corresponds to the limit of large 't Hooft coupling. Therefore
one can make predictions for the strongly coupled CFT using the
correspondence but these can generally only be checked for certain
protected objects.

More recently the opposite limit, in which the CFT is weakly
coupled, has been discussed by a number of
people~\cite{smallradius}. In~\cite{bbms} the string spectrum in an
$AdS_5 \times S^5$ background at small radius was extrapolated and
it precisely matches the operator spectrum of free $\cal N$= 4 SYM
in the planar limit. In particular the limit of zero YM coupling has
been conjectured to be dual to a massless higher spin field theory
which, although inconsistent when coupled to gravity in flat
space-time, can be consistently defined in AdS
spaces~\cite{Vasiliev90} (for a review see~\cite{vasiliev} and
references therein). Turning on the coupling in the YM side
corresponds in AdS to a Higgs mechanism, in which the massless
higher spin fields develop a mass, essentially by eating lower spin
Goldstone fields. This phenomenon was termed `La Grande Bouffe'
in~\cite{bms}. The remaining massless fields will all be contained
in the supergravity multiplet. In the dual CFT at zero coupling
there are infinitely many higher spin conserved currents (in
one-to-one correspondence with the AdS higher spin gauge fields).
The CFT counter-part of `La Grande Bouffe' is thus the anomalous
violation of these conserved currents when the coupling is turned
on, with the only remaining conserved currents lying in the energy
momentum tensor multiplet.

The simplest example of a (bosonic) higher spin $s$ field is that of
a tensor with $s$ completely symmetrised spacetime indices. For such
an object, in flat spacetime, the massless limit of a massive spin
$s$ field gives rise to $s+1$ massless fields of spins $0,1,\dots
s$. However the AdS/CFT correspondence predicts that the massless
limit of a massive spin $s$ field in AdS is a massless spin $s$
field  and a massive spin $s-1$ field. The reason for this is that
HS currents $J_{i_1 \dots i_s}$ with $s
>2$ occur in ${\cal N}=4$ SYM, where they are conserved at vanishing
coupling $g=0$, and conformal invariance fixes the dimension of such
a spin $s$ conserved current on the $d$ dimensional boundary to be
$s+d-2$. Interactions are responsible for their anomalous violation
$$ \de^{i_1} J_{i_1 \dots i_s} = g {\cal X}_{i_2 \dots
i_s}\label{anom} \quad ,
$$
and in the zero coupling limit the dimension of ${\cal X}$ is
$s+d-1$. This implies that ${\cal X}$ is not a conserved spin $s-1$
current when $g=0$, and therefore one expects it to be dual to a
massive field in the bulk.

In a recent paper~\cite{lagrandebouffe} the St\"uckelberg
formulation of bosonic massive higher spin fields (with completely
symmetrised spacetime indices) in AdS was derived (see
also~\cite{DESWALD,ZINO} for similar results). These massive
equations in AdS were then used to extrapolate the massless limit,
and one indeed obtains a massless spin $s$ field and a massive spin
$s-1$ field in line with the CFT predictions (this phenomenon has
also been discussed from a cosmological viewpoint in~\cite{DESWALD}
where it was termed `partial masslessness'). The mass of the spin
$s-1$ field one obtains in this way is precisely the one predicted
by AdS/CFT.

Here we first review the results of~\cite{lagrandebouffe}. We will
only focus on higher spin bosons whose spacetime indices are
completely symmetrised, although the fermionic case has been
analysed in~\cite{fermionicbouffe} and turns out to reveal similar
features. We then conclude with some comments on how to generalize
our analysis to other higher spin representations and some
speculations on how to trigger {\it la Grande Bouffe} in the AdS
bulk.

\section{Higgs \'a la St\"uckelberg for higher spin fields: flat space vs AdS}

An easy way to derive the St\"uckelberg formulation of a massive
spin $s$ field in flat $D$ dimensional spacetime is to consider a
massless spin $s$ field in $D+1$ dimensions~\cite{tesifabio},
described in terms of a symmetric rank $s$
tensor\footnote{Parentheses $(...)$ denote symmetrization of
spacetime indices with strength one. This is the origin of factors
of $s$ in several formulae.} $\Phi_{M_1...M_s}$ ($M_i$ are
$D+1$-dimensional spacetime indices) satisfying the condition \be
\Phi^L{}_L{}^M{}_M{}_{M_5 ... M_s}=0 \label{doubletrace} \quad . \ee
The resulting equation is invariant with respect to the gauge
transformation
  \be
  \delta\Phi_{M_1...M_s}=s\de_{(M_1}\e_{M_2...M_s )}\quad ,
  \label{delta5}
  \ee
where the gauge parameter $\e$ is symmetric and
traceless~\cite{FRONSDAL}. In order to perform a `massive' KK
dimensional reduction, one has to consider the various component
fields to depend harmonically on the extra coordinate $y$,
  \be \Phi_{\m_1...\m_{s-k}y...y} (x,y)=
  (i)^k\phi^{(s-k)}_{\m_1...\m_{s-k}}(x) e^{imy} + {\rm c.c.}\quad .
  \label{realfields}
  \ee
By taking linear combinations one can choose the fields
$\phi^{(s-k)}$ to be real in $D$ dimensions. The gauge
transformation (\ref{delta5}) becomes for the D-dimensional fields
  \be \delta\phi^{(s-k)}_{\m_1...\m_{s-k}}=(s-k)\de_{(\m_1}
  \e^{(s-k-1)}_{\m_2...\m_{s-k})}+ k
  m\e^{(s-k)}_{\m_1...\m_{s-k}}\quad, \label{fourpointeight}
  \ee
where the D-dimensional gauge parameters related to the
(D+1)-dimensional ones by means of
  \be
  \e_{\m_1...\m_{s-k-1}y...y}
  (x,y)=(i)^k\e^{(s-k-1)}_{\m_1...\m_{s-k-1}} e^{imy} + {\rm c.c.}
  \label{doubletraceps}\quad.
  \ee

If $m \neq 0$, from eq. (\ref{fourpointeight}) it turns that only
$\phi^{(s)}$ does not transform algebraically with respect to any
gauge transformations. Actually, not all the lower spin components
can be put to zero fixing their gauge invariance, because of the
traceless constraint on the gauge parameters. The remaining lower
spin fields, that are identically zero on shell, are the auxiliary
fields of the massive theory \cite{singhhagen}, and one ends up with
an equation for a massive spin $s$ field $\phi^{(s)}$. If $m =0$,
instead, none of the gauge parameters can be used to gauge away any
of the fields, and therefore all the fields $\phi^{(s-k)}$,
$k=0,1,...,s$, become massless.

We now want to consider the same system of equations in AdS. In
particular, we consider the field equation for $\phi^{(s-1)}$, and
we gauge away $\phi^{(s-2)}$ and $\phi^{(s-3)}$ using $\e^{(s-2)}$
and $\e^{(s-3)}$. This implies that only the fields $\phi^{(s)}$
and $\phi^{(s-1)}$ will appear in the equation, while all the
other fields $\phi^{(s-k)}$, with $k=4,...,s$ are auxiliary
fields. The only gauge invariance left is the one with respect to
the {\it traceless} gauge parameter $\e^{(s-1)}$, and in AdS it
will require the addition of a mass term for $\phi^{(s-1)}$. We
then go to the $m=0$ limit, and we check whether the mass term we
included leads to a gauge symmetry enhancement or not. Since we
don't find any inconsistency, that would indicate gauge symmetry
enhancement, we can continue the procedure to $m=0$ and we end up
with a massless spin $s$ and a massive spin $s-1$.

Before we proceed, we first observe that gauge invariance (\ie
masslessness) in AdS implies the presence of a mass-like term in the
field equations, proportional to the inverse of the AdS radius $L$.
In particular, the equation for a spin $l$ field
  \bea & &  \Box \phi^{(l)}_{\m_1 ... \m_l} - l
  \nabla_{(\m_1} (\nabla \cdot \phi^{(l)} )_{\m_2 ... \m_l) } +
  \frac{l(l-1)}{2} \nabla_{(\m_1}
  \nabla_{\m_2} \phi^{(l)\l}{}_{\l \m_3 ... \m_l)} \nonumber \\
  & & - M_{AdS}^2 \ \phi^{(l)}_{\m_1 ... \m_l} - \tilde{M}_{AdS}^2 \
  g_{(\m_1\m_2} \phi^{(l)\l}{}_{\l \m_3 ... \m_l)}=0
  \label{masslessspinsads} \quad ,
  \eea
with
  \be M_{AdS}^2 = \frac{(l-2)(D-1) + (l-1)(l-4)}{L^2} \quad ,
  \quad \tilde{M}_{AdS}^2 = \frac{l(l-1)}{L^2} \label{adsmasses}\quad
  ,
  \ee
is gauge invariant with respect to
  \be \d \phi^{(l)}_{\m_1... \m_l}
  = l \nabla_{(\m_1} \e^{(l-1)}_{\m_2 ... \m_l)} \quad ,
  \ee
where $\e$ is traceless.

We now consider in AdS the equation for the spin $s-1$
St\"uckelberg field $\phi^{(s-1)}$, that we denote here by
$\chi^{(s-1)}$ for clarity. Requiring that this equation is
invariant with respect to the gauge transformations
  \be \d \phi^{(s)}_{\m_1
  ...\m_s} = s \nabla_{(\m_1} \e_{\m_2 ...\m_s)} \quad , \quad \d
  \chi^{(s-1)}_{\m_1 ...\m_{s-1}} = m \e_{\m_1 ...\m_{s-1}} \quad ,
  \label{gaugephichi}
  \ee
with $\e$ traceless, implies the presence of a mass proportional to
the inverse of the AdS radius, so that one ends up with the equation
\bea & & \Box \chi^{(s-1)}_{\m_1 ...\m_{s-1}} - (s-1) \nabla_{(\m_1}
( \nabla \cdot
\chi^{(s-1)})_{\m_2 ... \m_{s-1})} \nonumber \\
& & +\frac{(s-1)(s-2)}{2}\nabla_{(\m_1}
\nabla_{\m_2} \chi^{(s-1)\l}{}_{\l\m_3 ... \m_{s-1})} \nonumber \\
& & - m (\nabla \cdot \phi^{(s)} )_{\m_1 ... \m_{s-1}} + (s-1) m
\nabla_{( \m_1} \phi^{(s)\l}{}_{\l\m_2 ...\m_{s-1})} \nonumber \\
& &- \frac{(s-1) [ (D-1) + (s -2)]}{L^2} \chi^{(s-1)}_{\m_1
...\m_{s-1}}=0 \quad .\label{chieqmassads}\eea We thus would like
to compare this mass term with the first of eqs.
(\ref{adsmasses}), where $l=s-1$. They are definitely different,
which means that no symmetry enhancement occurs when $m=0$, and
any massive spin $s$ field in the limit of zero mass decomposes
into a massless spin $s$ field and a massive spin $s-1$ field. In
other words, the new feature of AdS is the fact that the auxiliary
field structure is preserved for the spin $s-1$ field even when
the spin $s$ field becomes massless. Note that our procedure
leaves undetermined a possible mass term of the form $1/L^2
g_{(\m_1\m_2} \chi^{(s-1)\l}{}_{\l\m_3 ...\m_s)}$ in eq.
(\ref{chieqmassads}), since $\chi^{(s-1)\l}{}_\l$ is gauge
invariant with respect to (\ref{gaugephichi}). This is not an
issue as long as we focus on the field equations, since
$\chi^{(s-1)\l}{}_\l$ can be put to zero on shell using the lower
rank equations. Nevertheless, the whole set of equations can be
derived from a lagrangian once the correct equations for the
auxiliary fields are introduced, in a similar way to the flat
space case.

The difference between the AdS mass term and this mass term (for
simplicity we define $s' =s-1$ from now on) is \be - \frac{2(D-1) +4
(s'-1)}{L^2} \chi^{(s')}_{(\m_1 ...\m_s')} \quad . \ee We therefore
get \be M^2 L^2 = 2(D-1) +4 (s'-1) \quad . \ee In $D=5$ ($d=4$) this
equation becomes \be M^2 L^2 = 4(s'+1) \quad . \ee This is exactly
what we get from the standard relation between mass in AdS and
dimension of the dual operator in the boundary theory
\cite{ferrara,bms}, \be M^2 L^2 = \Delta (\Delta - 4 ) -
\Delta_{min} (\Delta_{min}-4) \quad , \label{identity}\ee with \be
\Delta = s'+ 4 \quad , \qquad \Delta_{min}= s'+2 \quad , \ee which
is exactly the dimension of the corresponding spin $s' $ operator at
vanishing Yang-Mills coupling. For arbitrary dimension $d = D-1$,
$\Delta_{min} = s' + d -2$ represents the unitary bound for the
dimension of a spin $s'$ current, \ie a totally symmetric rank
$s'$ classically conserved tensor current, and the identity
(\ref{identity}), with 4 substituted with $d$, is satisfied with
$\Delta = s'+d$.

The case of an anomalous spin $s=1$ (axial) vector current, for
which the relevant St\"uckelberg field in the bulk is a massless
(pseudo) scalar, is special. Indeed, eqs.
(\ref{masslessspinsads},\ref{adsmasses}) are meaningless for a
scalar field, while eq. (\ref{chieqmassads}) shows that for $s=1$
the mass for $\chi^{(0)}$ vanishes. This agrees with the
mass/dimension relation for scalar fields,
  \be M^2 L^2 = \Delta
  (\Delta - d ) \quad , \ee
since $\chi^{(0)}$ is dual to a naively marginal scalar operator of
dimension $\Delta = d = D-1$.

\section{Conclusions and perspectives}
We would like to conclude with some comments on how to generalize
our analysis to other higher spin representations and some
speculations on how to trigger {\it la Grande Bouffe} in the AdS
bulk.

In the introduction we already mentioned how the case of a
spontaneously broken fermionic spin $s+1/2$ symmetry in AdS can be
described \cite{fermionicbouffe} along the lines of the bosonic spin
$s$ totally symmetric tensors \cite{lagrandebouffe} reviewed above.
The two cases can indeed be related by exploiting the global
supersymmetries present in AdS even after HS symmetry breaking. By
the same token one should be able to relate more general HS bosonic
and fermionic representations that appear in the AdS HS
supermultiplets for $D=5$ and higher. Indeed the analysis of
\cite{mbjfmhs,nbmbjfmhs1,nbmbjfmhs2} shows that all the relevant
Goldstone modes are present in the free SYM spectrum that are needed
to achieve HS breaking in a way compatible with full $PSU(2,2|4)$
symmetry. In particular, it is shown there that the string spectrum
on $AdS_5\times S^5$ can be extrapolated to the HS enhancement point
where it can be precisely matched with the spectrum of single trace
gauge invariant operators and decomposed into representations of
$HS(2,2|4)$. The latter, indicated as {\it YT-pletons}, are in one
to one correspondence with the Yang Tableaux (YT) compatible with
the cyclicity of the trace over the colour indices which is the
counterpart of `level matching', \ie the closure of the string. To
wit, the HS gauge fields belong to the {\it doubleton} $\Phi_2$
corresponding to the bilinear gauge invariant operators ${\cal
O}^{AB}_2 = tr(W^A W^B)$, where $W^A$ denotes any of the `letters'
of the SYM `alphabet', \ie any of the fundamental SYM fields or
derivatives thereof, modulo field equations. Similarly there are two
{\it tripletons}: the $f-{\it tripleton}$ ${\cal X}_3$,
corresponding to the totally antisymmetric trilinear operators
${\cal O}^{[ABC]}_{3, f} = tr(W^A [W^B, W^C])$, expressible  in
terms of the structure constants $f_{abc}$, which contains the first
generation of Goldstone modes, and the $d-{\it tripleton}$
${\Phi}_3$, corresponding to the totally symmetric trilinear
operators ${\cal O}^{(ABC)}_{3,d} = tr(W^A \{ W^B, W^C\})$,
expressible in terms of the cubic Casimir $d_{abc}$, which accounts
for the first KK recurrence of the supergravity fields and their HS
cousins. No {\it hooked} YT with three boxes are compatible with
gauge symmetry, \ie cyclicity of the trace or equivalently level
matching / closure of the string. By the same token no two-particle
{\it tripletons} are allowed that would correspond to double-trace
trilinear operators. Double-trace operators first appear in {\it
tetrapletons}, \eg $\Phi_{2,2}=\Phi_2\Phi_2$. Other {\it
tetrapletons} can be of various kinds. The $q-{\it tetrapleton}$
${\Phi}_4$, corresponding to the totally symmetric quadrilinear
operators ${\cal O}^{(ABCD)}_{4,q} = tr(\{W^A, W^{(B}\} \{ W^{C)},
W^D\})$, expressible in terms of the quartic Casimir $q_{abcd}=
\delta^{ef}d_{e(ab}d_{cd)f}$, which accounts for the second KK
recurrence of the supergravity fields and their HS cousins. The
$w-{\it tetrapleton}$ ($w$ stands for window: $w \approx ff$) ${\cal
X}_4$, corresponding to the quadrilinear operators ${\cal
O}^{[AB][CD]}_{4,w} = tr([W^A, W^B][W^C, W^D])$, expressible in
terms of $w_{[ab][cd]} = \delta^{ef}f_{abe}f_{cdf}$, which accounts
for the second generation of Goldstone modes. Moreover, the $h-{\it
tetrapleton}$ ($h$ stands for hooked) ${\Psi}_4$, corresponding to
the mixed symmetry quadrilinear operators ${\cal O}^{(A[B)CD]}_{4,h}
= tr(\{W^A, W^{B}\}  [ W^{C)}, W^D])$, expressible in terms of
$h_{(a[b)cd]}= \delta^{ef}d_{ea[b}f_{cd]f}$, which accounts for
genuinely massive HS fields.

We are now ready to speculate on the structure of the Lagrangian, if
any, that governs the dynamics of the massless and massive HS
multiplets at the point of HS enhancement. We heavily rely on the
known properties of free SYM theory that should provide the
holographic dual. We start by choosing a normalization such that
two-point functions are normalized to 1 and focus only on the master
fields\footnote{We are using the term master field in a loose sense.
Strictly speaking the massless {\it doubleton} $\Phi_2$ requires two
master fields in Vasiliev description~\cite{Vasiliev90,vasiliev}, a
master connection ${\cal W}$ and a master scalar curvature ${\cal
R}$. For massive HS multiplets such as ${\cal X}_3$ and ${\cal X}_4$
no explicit master field description is available at present.}
$\Phi_2$, ${\cal X}_3$, ${\cal X}_4$, then schematically
  \bea
  {\cal L}_{HS} &=& \Phi_2
  \Phi_2 + {1\over N} \Phi_2 \Phi_2 \Phi_2 + {1\over N^2} \Phi_2
  \Phi_2 \Phi_2 \Phi_2+ ...
  \\
  & & {\cal X}_3 {\cal X}_3 + {1\over N} {\cal X}_3 {\cal X}_3 \Phi_2
  + {1\over N^2} ({\cal X}_3{\cal X}_3 \Phi_2 \Phi_2 + {\cal X}_3{\cal
  X}_3 {\cal X}_3{\cal X}_3) + ...
  \nonumber \\
  & & {\cal X}_4 {\cal X}_4 + {1\over N} ({\cal X}_4 {\cal X}_3 {\cal
  X}_3   + {\cal X}_4 \Phi_2 \Phi_2 ) + {1\over N^2} ({\cal X}_4 {\cal
  X}_3 {\cal X}_3 \Phi_2 + {\cal X}_4 \Phi_2 \Phi_2 \Phi_2 + {\cal
  X}_4{\cal X}_4 \Phi_2 \Phi_2+ \nonumber \\
  && {\cal X}_4{\cal X}_4 {\cal X}_3{\cal X}_3) + {\cal X}_4{\cal
  X}_4{\cal X}_4{\cal X}_4) + ... \nonumber
  \eea
Notice that any term in ${\cal L}_{HS}$ encompasses an infinite
number of terms / couplings involving the component HS fields as
well as their derivatives to arbitrary high order. The schematic
form of ${\cal L}_{HS}$ seems already to be in conflict with the
consistent truncation to the massless HS theory that only involves
$\Phi_2$ \cite{sezsund}, and sets ${\cal X}_3$, ${\cal X}_4$ to
zero. Notice however that the offending coupling ${\cal X}_4 \Phi_2
\Phi_2$ is extremal and as such, in the spirit of the holographic
correspondence, is expected to arise from a purely boundary
contribution since the relevant bulk integral would diverge (at
least for the lowest scalar components and the rest should follow
from $HS(2,2|4)$ if not $PSU(2,2|4)$ symmetry). Alternatively one
can define ${\cal X}_4$ in such a way that it is `orthogonal' to
$\Phi_{2,2}= \Phi_2 \Phi_2$. From now on we assume that this is the
case. It is also convenient to rescale the master fields $\Phi_2$,
${\cal X}_3$, ${\cal X}_4$ by a factor of $N$ so that ${\cal
L}_{HS}$ displays an overall factor of $N^2$ (that should in the end
become an  $N^2-1$ for $SU(N)$) \bea {\cal L}_{HS} &=& N^2 (\Phi_2
\Phi_2 + \Phi_2 \Phi_2 \Phi_2 + \Phi_2 \Phi_2 \Phi_2 \Phi_2+ ...
\\
& & {\cal X}_3 {\cal X}_3 + {\cal X}_3 {\cal X}_3 \Phi_2 +  {\cal
X}_3{\cal X}_3 \Phi_2 \Phi_2 + {\cal X}_3{\cal X}_3 {\cal X}_3{\cal
X}_3 (1 + {\cal O}(1/N^2)) + ...
\nonumber \\
& & {\cal X}_4 {\cal X}_4 + {\cal X}_4 {\cal X}_3 {\cal X}_3  +
{\cal X}_4 \Phi_2 \Phi_2 + {\cal X}_4 {\cal X}_3 {\cal X}_3 \Phi_2 +
{\cal X}_4 \Phi_2 \Phi_2 \Phi_2 + {\cal X}_4{\cal X}_4 \Phi_2 \Phi_2
+
\nonumber \\
&&+ {\cal X}_4{\cal X}_4 {\cal X}_3{\cal X}_3 (1 + {\cal O}(1/N^2))
+ {\cal X}_4{\cal X}_4{\cal X}_4{\cal X}_4) (1 + {\cal O}(1/N^2))+
... \quad .\nonumber \eea As indicated non-planar corrections
(subleading in $1/N$) affect some of the couplings, \ie those
involving at least six $f$'s. Inclusion of the other {\it
YT-pletons} should not change the structure in any significant way.

Assuming that HS symmetry be strong enough to fix completely the
detailed structure of ${\cal L}_{HS}(\Phi_2, {\cal X}_3,{\cal X}_4,
...)$, we can take it as the starting point to discuss HS symmetry
breaking. Holography suggests there must be a massless (complex)
scalar, the (complexified) dilaton, dual to the SYM interaction
lagrangian ${\cal L}_{SYM}$, that could acquire a VEV preserving
exact $PSU(2,2|4)$, while breaking $HS(2,2|4)$. This
scalar\footnote{We here mean the component dual to the SYM coupling
$g_{YM}$, which is known to be exactly marginal. We neglect for the
time being $\vartheta_{YM}$ that is also exactly marginal but can
only show up in non-perturbative corrections induced by
(D)-instantons\cite{mbmgskgcr} .} is a combination of the the
massless scalar singlets that appear in $\Phi_2, {\cal X}_3$ and
${\cal X}_4$. Once it takes a VEV, all but a handful of HS gauge
fields in $\Phi_2$ become massive by eating lower spin Goldstone
modes in ${\cal X}_3$ which in turn absorb the second generation
Goldstone modes in ${\cal X}_4$. In order to make contact with the
discussion of the totally symmetric bosonic spin $s$ tensors, the
parameter $M$ there should be thought of as the relevant VEV.

Admittedly the resulting picture is still obscure due to the lack
of knowledge of an efficient description of massive HS multiplets
\'a la Vasiliev. Yet it is quite appealing how subtle effects such
as the anomalous violation of HS currents might be described in
rather compact terms as hinted at above. To be concrete, one may
try to use HS symmetry, that acts linearly on the master fields
prior to its spontaneous breaking, to fix the relative strength of
some of the trilinear couplings that are responsible for the
spontaneous breaking and the consequent mass generation once the
`dilaton' gets a VEV. This is our main goal in the near future.

{}{}
{}
{}

\section*{Acknowledgments}
We are grateful to P. Heslop for a very stimulating collaboration.
F.R. would like to thank INFN for support and the Department of
Physics for hospitality at University of Rome `Tor Vergata' where
this work has been completed. The work of M.B. was supported in
part by INFN, by the MIUR-COFIN contract 2003-023852, by the EU
contracts MRTN-CT-2004-503369 and MRTN-CT-2004-512194, by the
INTAS contract 03-516346 and by the NATO grant PST.CLG.978785. The
work of F.R. is supported by a European Commission Marie Curie
Postdoctoral Fellowship, Contract MEIF-CT-2003-500308.

\end{document}